\documentclass[11pt]{cernrep}

\usepackage{graphicx}

\usepackage{amsmath}

\begin{document}

 \title{ANGULAR DISTRIBUTION OF DRELL YAN PAIRS IN p+A AT LHC}

\author{R.~J.~Fries}

\institute{Physics Department, Duke University, P.O.Box 90305, Durham, 
  NC 27708}

\maketitle

\begin{abstract}
We discuss the angular distribution of dilepton pairs 
in p+A collisions at LHC at low  momentum transfer and 
transverse momentum. This includes helicity amplitudes and 
angular coefficients as well as a look at the Lam-Tung sum rule. We use 
leading twist NLO calculations and higher twist corrections by double 
scattering in large nuclei.
\end{abstract}

\section{INTRODUCTION}

The Drell Yan cross section can be described by a contraction of the lepton 
and hadron tensors
\begin{equation}
  \frac{d\sigma}{d Q^2 \, d q_\perp^2 \, d y \, d \Omega} = 
  \frac{\alpha^2}{64 \pi^3 SQ^4}
  L_{\mu\nu}W^{\mu\nu}.
\end{equation}
Here we parametrize the lepton pair by
the invariant mass $Q$ of the virtual photon and its transverse momentum 
$q_\perp$ and rapidity $y$ in the center of mass frame of the colliding
hadrons. In addition we give the direction of one 
of the leptons, say the positively charged one, in a photon rest frame 
using polar and azimuthal angles $\phi$ and $\theta$: $d\Omega=d\phi \, d\cos
\theta$.
Now the cross section can be understood in terms of four helicity amplitudes
\cite{Collins:iv,Lam:pu,Mirkes:1992hu}
\begin{equation}
  \begin{split}
  \frac{d\sigma}{d Q^2 \, d q_\perp^2 \, d y \, d \Omega} = 
  \frac{\alpha^2}{64 \pi^3 SQ^2} \Big[ W_{\rm TL} (1+ \cos^2 \theta) +
  W_{\rm L} (1/2 - 3/2 \cos^2 \theta)  \\ + W_{\Delta} \sin 2\theta \cos\phi
  + W_{\Delta\Delta} \sin^2 \theta \cos 2\phi \Big].
  \end {split}
  \label{eq:dy}
\end{equation}
These are defined as contractions $W_{\sigma,\sigma'} = \epsilon_\mu(\sigma)
W^{\mu\nu} \epsilon^*_\nu(\sigma')$
of the hadron tensor with polarization 
vectors of the virtual photon for polarizations $\sigma=0,\pm 1$. Only four
out of all possible contractions are independent, the others can be related 
by symmetries of the hadron tensor. The usual choice is to 
pick the longitudinal $W_{\rm L}= W_{0,0}$, the 
helicity flip $W_{\Delta} = (W_{1,0} + W_{0,1})/\sqrt{2}$ and the double 
helicity flip
amplitude $W_{\Delta\Delta} = W_{1,-1}$ together with the trace
$W_{\rm TL} = W_{T}+W_{L}/2=-W^\mu_\mu /2$ as a basis.
Note that integration over the angles $\theta$ and $\phi$ leaves only 
contributions from the trace
\begin{equation}
  \frac{d\sigma}{d Q^2 \, d q_\perp^2 \, d y} = \frac{\alpha^2}{64 \pi^3 SQ^2}
  \frac{16 \pi}{3} W_{\rm TL} =
  \frac{\alpha^2}{24 \pi^2 SQ^2} (-g_{\mu\nu}) W^{\mu\nu}.
\end{equation}

On the other hand, if we are only interested in relative angular 
distributions --- i.e.\ in the ratio 
\begin{equation}
  \frac{16\pi}{3} \> \frac{d\sigma}{dQ^2 \, 
  dq_\perp^2 \, dy \, d\Omega}\> \Big/ \>
  \frac{d\sigma}{dQ^2 \, dq_\perp^2 \, dy }
\end{equation}
--- we can make use of angular coefficients. Two different sets can 
be found in the literature \cite{Fries:2000da}. One set consists of the 
coefficients
\begin{equation}
  \label{eq:angcoeff}
  A_0 = \frac{W_{\rm L}}{W_{\rm TL}}, \quad A_1 = 
  \frac{W_{\Delta}}{W_{\rm TL}}, \quad
  A_2 = \frac{2 W_{\Delta\Delta}}{W_{\rm TL}},
\end{equation}
the other one is defined by
\begin{equation}
  \lambda = \frac{2-3A_0}{2+A_0}, \quad \mu = \frac{2A_1}{2+A_0}, 
  \quad \nu = \frac{2A_2}{2+A_0}.
  \label{eq:angcoeff2}
\end{equation}

\begin{figure}
\begin{center}
\includegraphics[width=7.5cm]{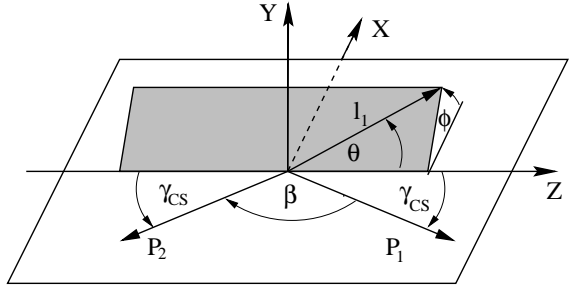} 
\caption{The Collins-Soper frame: the $z$-axis cuts the angle between
 $\mathbf{P}_1$ and $-\mathbf{P}_2$ into halves (the half angle is called the 
 Collins-Soper angle $\gamma_{\rm CS}$) while the $x$-axis is 
 perpendicular to $\mathbf{P}_1$ and $\mathbf{P}_2$. The direction of one
 lepton momentum $\mathbf{l}_1$ can then be given by the angles $\theta$ and 
 $\phi$ }
\label{fig:csframe}
\end{center}
\end{figure}

The helicity amplitudes are frame dependent. In principle we allow all frames
where the photon is at rest, i.e.\ $q^\mu = (Q,0,0,0)$. However
there are some frames with particular properties studied in \cite{Lam:pu}. 
Here we only use the Collins-Soper (CS) frame. 
It is characterized by two properties. First the $y$-axis is perpendicular to 
the plane spanned by the two hadron momenta $\mathbf{P}_1$ and $\mathbf{P}_2$ 
(which are no longer collinear in a photon rest frame as long as $q_\perp 
\not= 0$, what is true in our kinematic domain) and second the $z$-axis cuts 
the angle between $\mathbf{P}_1$ and $-\mathbf{P}_2$ into two equal halves, see
Fig.~\ref{fig:csframe}.

\section{LEADING TWIST}

\begin{figure}[b]
\begin{center}
\includegraphics[width=5cm]{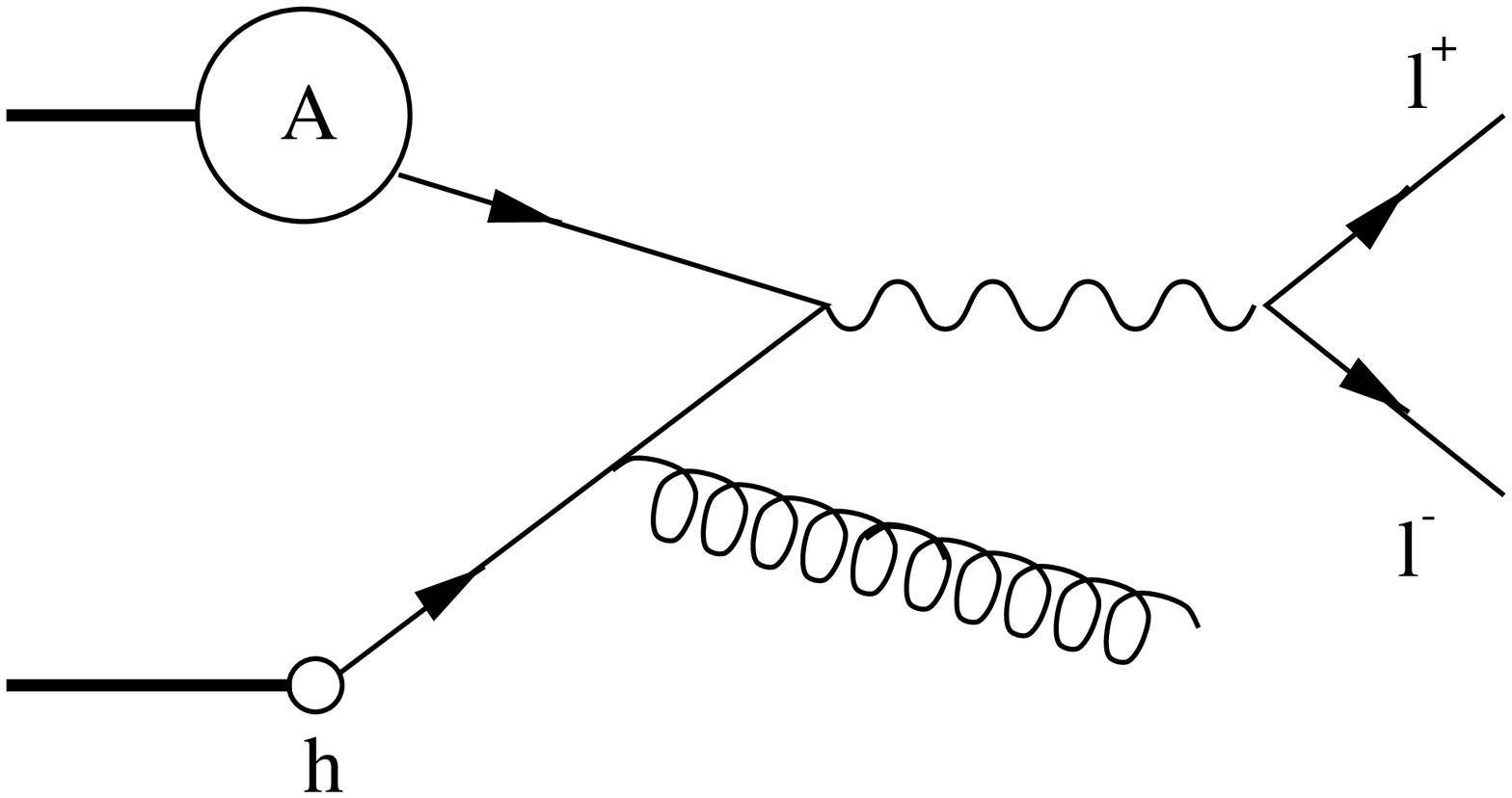} \hspace{4em}
\includegraphics[width=5cm]{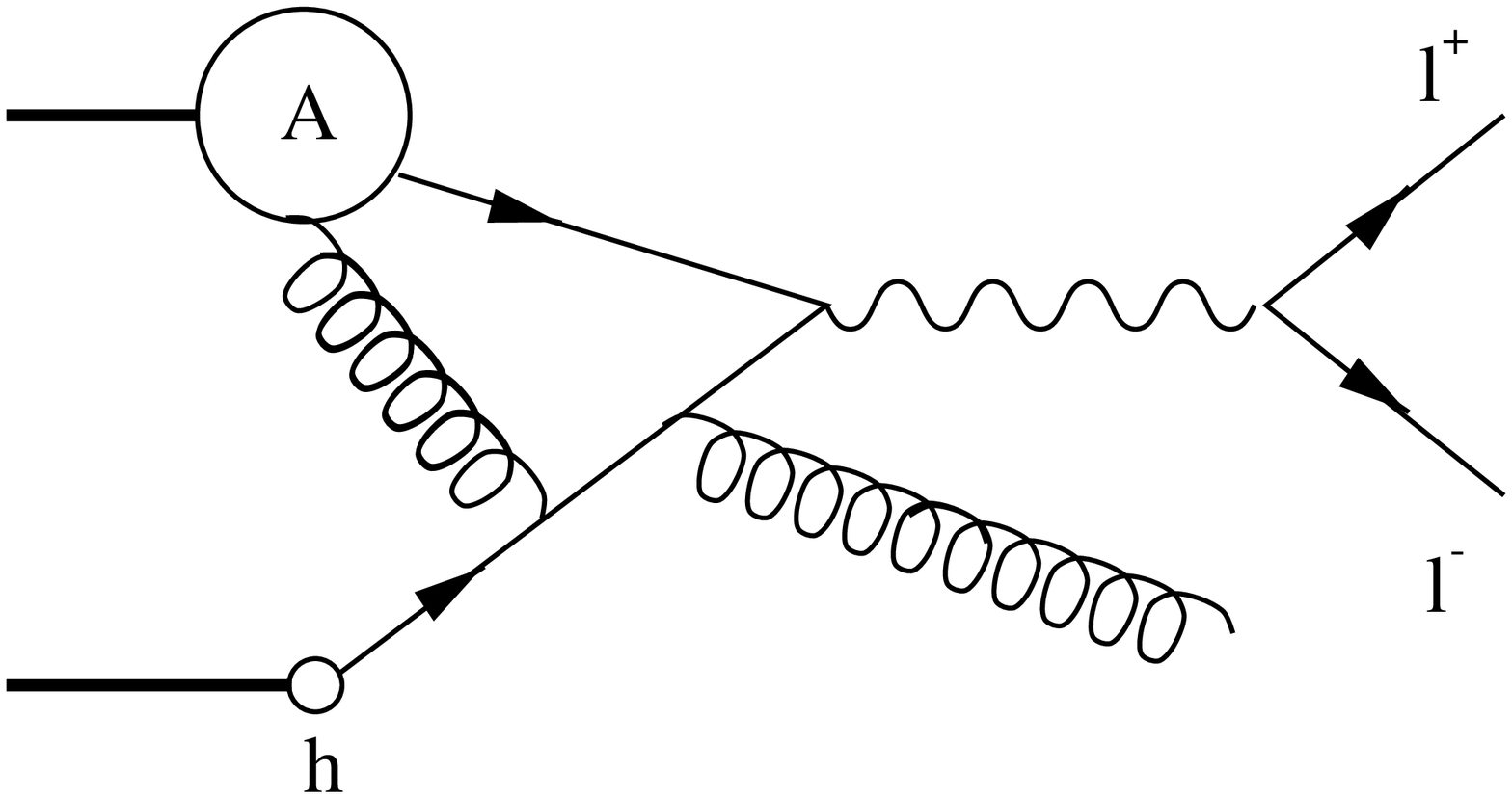}
\caption{Examples for diagrams contributing to the dilepton production in
hadron (h) nucleus (A) scattering at 
twist-2 (left) and twist-4 (right) level}
\label{fig:diag}
\end{center}
\end{figure}

The hadron tensor for the Drell Yan process in a leading twist (twist-2)
calculation is given by the well known factorization formula as a convolution
of two parton distributions with a perturbative parton cross section.
We are interested here in the kinematic region characterized by intermediate
photon mass $Q$ and intermediate transverse momentum $q_\perp \sim Q$ of a 
few GeV. For these values the dominant contribution to twist-2 is given
by the next-to-leading order (NLO) perturbative diagrams like the one in 
Fig.~\ref{fig:diag} (left) and we can safely omit logarithmic corrections of 
type $\ln^2 Q^2/q_\perp^2$. It has been shown that a leading twist
calculation up to NLO respects the so called Lam Tung sum rule 
$W_L = 2 W_{\Delta\Delta}$ \cite{Lam:1980uc}.
In terms of angular coefficients this can be rewritten as $A_0=A_2$ or
$2\nu=1-\lambda$. Furthermore the spin flip amplitude $W_\Delta$ has to vanish
for a symmetric colliding system like $p+p$. 
For $p+A$ we expect small contributions for $W_\Delta$ 
due to lost isospin symmetry and nuclear corrections to the parton 
distributions. Results for $p+p$ at $\sqrt{S}=5.5$ TeV have already been 
presented elsewhere \cite{Gavin:1995ch}.

\section{NUCLEAR ENHANCED TWIST-4}

For large nuclei corrections to the leading twist calculation, induced by 
multiple scattering, play an important role.
The formalism how to take into account these nuclear enhanced
higher twist contributions was worked out by Luo, Qiu and 
Sterman \cite{Luo:fz,Luo:ui,Luo:np}. The leading nuclear corrections 
(twist-4 or double scattering) have already been calculated for some 
observables. For Drell Yan this was first done by Guo \cite{Guo:1998rd} and 
later generalized to the Drell Yan angular distribution 
\cite{Fries:2000da,Fries:1999jj}.

Fig.~\ref{fig:diag} (right) shows an example for a diagram contributing at 
twist-4 level. Now two partons $a$, $b$ from the nucleus and one ($c$) from 
the single hadron are involved. As long as $q_\perp \sim Q$ twist-4 is 
dominated by two different contributions. The double hard (DH) process where 
each parton from the nucleus has a finite momentum fraction and the soft hard 
(SH) process where one parton has vanishing momentum fraction. The 
factorization formulas are given by
\begin{eqnarray}
  \label{eq:twist-4}
  W^{\mu\nu} &=& \sum_{a,b,c} \int \frac{d x_c}{x_c} \> T^{\rm DH}_{ab}
  (x_a,x_b) \,
  H_{ab+c}^{\mu\nu} (q,x_a,x_b,x_c) \, f_c(x_c), \\
  W^{\mu\nu} &=& \sum_{a,b,c} \int \frac{d x_c}{x_c} \> D_{x_a,x_b}
  (q,x_c) T^{\rm SH}_{ab} (x_a) \,
  H_{ab+c}^{\mu\nu} (q,x_a,x_b=0,x_c) \, f_c(x_c)
\end{eqnarray}
for double hard and soft hard scattering respectively \cite{Fries:2000da,
Guo:1998rd,Fries:1999jj}. 
$D_{x_a,x_b}(q,x_c)$ is a second order differential operator in $x_a$ and 
$x_b$. 
$T^{\rm DH}_{ab}$ and 
$T^{\rm SH}_{ab}$ are new matrix elements of twist-4 which encode non 
perturbative correlations between the partons $a$ and $b$. Since we are still
missing solid experimental information about these new quantities, they are 
usually modeled in a simple way through parton distributions.
We use $T^{\rm DH}_{ab}(x_a,x_b)= C A^{4/3}f_a(x_a) f_b(x_b)$ and
$T^{\rm SH}_{ab}(x_a)= \lambda^2 A^{4/3}f_a(x_a)$ where $C$ and $\lambda^2$
are normalization constants. The key feature,
their nuclear enhancement, is their scaling with the nuclear size.

\begin{figure}[t]
\begin{center}
\includegraphics[width=7.0cm]{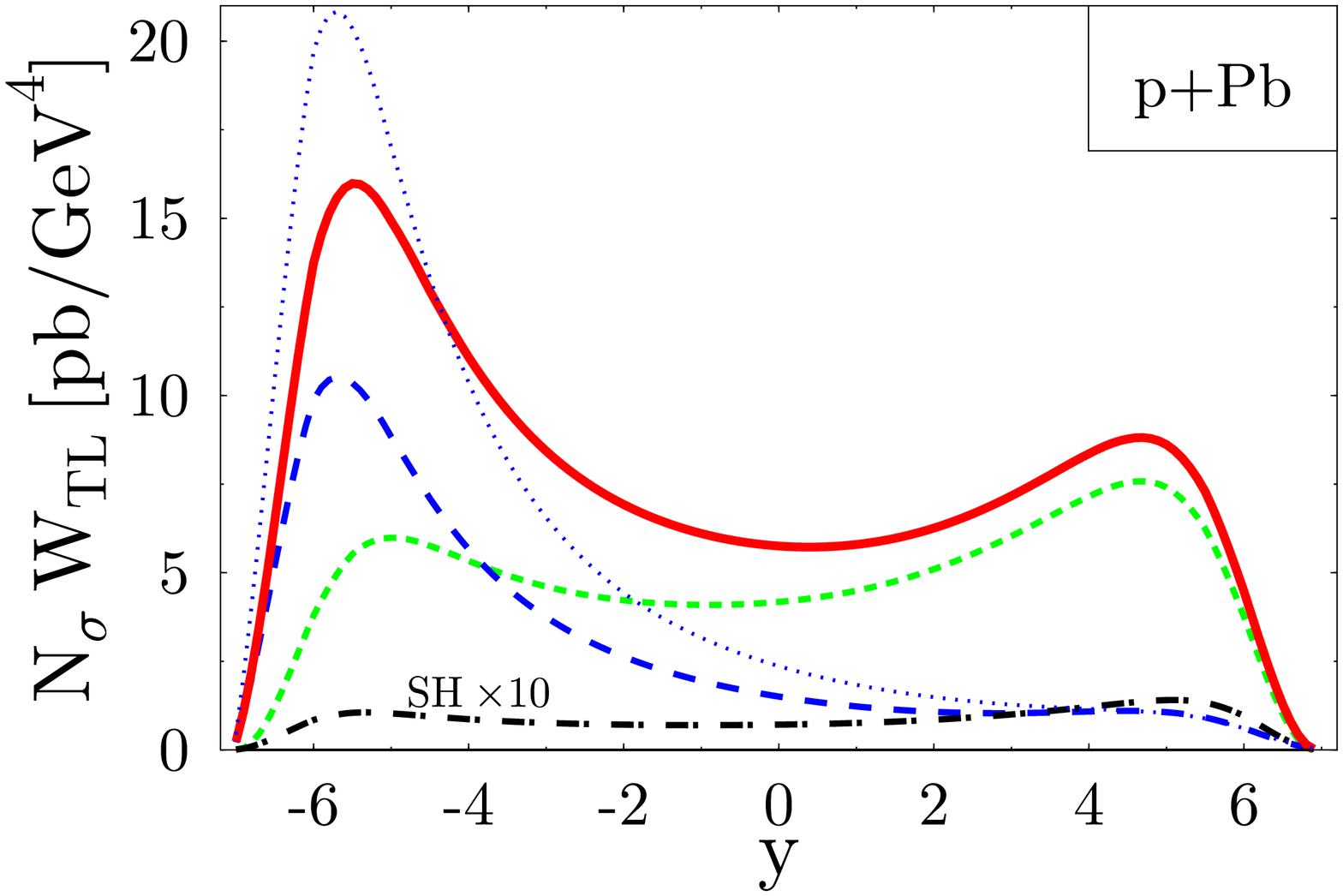} \hfill
\includegraphics[width=7.0cm]{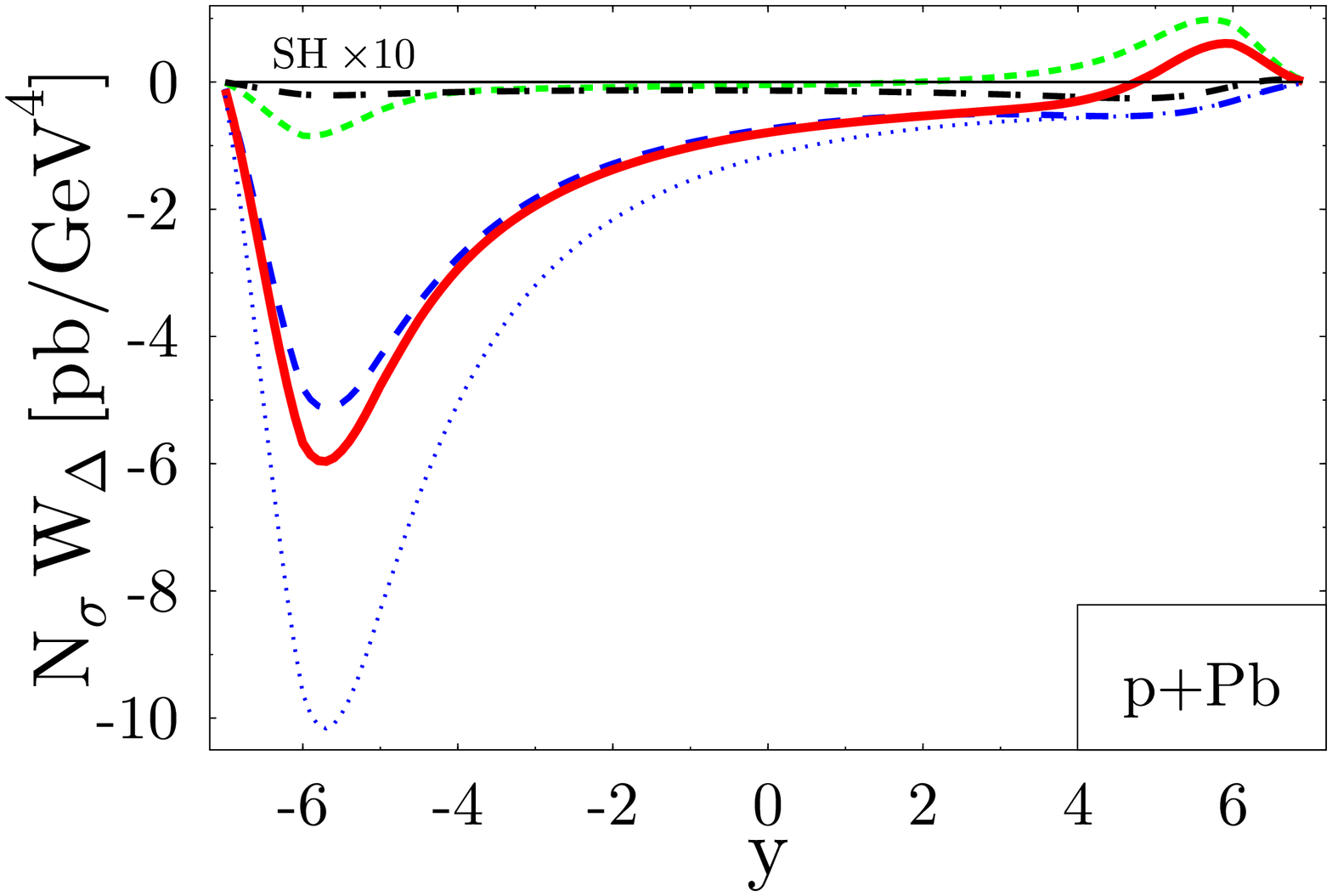}
\caption{Rapidity dependence of $N_\sigma W_{\rm TL}$ (left) and $N_\sigma 
 W_{\Delta}$ (right) for $p+Pb$ at $\sqrt{S}=8.8$ TeV, $Q=5$ GeV and 
 $q_\perp=4$ GeV: twist-2 NLO (short 
 dashed), double hard with EKS98 modifications (long dashed), soft hard
 (dash dotted, scaled up by a factor of 10) and the sum of all contributions
 (bold solid line). The double hard contribution 
 calculated without EKS98 modifications is also shown (dotted line). 
 Note that the incoming nucleus has positive rapidity.}
 \label{fig:wy}
\end{center}
\end{figure}

It has been shown in \cite{Fries:2000da,Fries:1999jj} that the DH process 
shows a trivial angular 
pattern in the sense that it is similar to the lowest order simple annihilation
of on shell quarks with the only difference that one of the quarks now carries
finite transverse momentum $q_\perp$. In this spirit it is no surprise that the
DH contribution respects the Lam Tung relation. On the other hand SH scattering
is more complicated and violates the Lam Tung sum rule.
Also we expect that the spin flip amplitude $W_\Delta$ can receive large
contributions from the twist-4 calculation.

\section{NUMERICAL RESULTS}

In this section we present some numerical results obtained for proton lead
collisions at the LHC energy $\sqrt{S}= 8.8$ TeV. Results for RHIC energies
can be found elsewhere \cite{Fries:2000da}.
We use CTEQ5L parton distributions \cite{Lai:1999wy} combined with EKS98
\cite{Eskola:1998df} nuclear 
nuclear modifications both for the nuclear parton distributions and for the 
models of the twist-4 matrix elements. In some plots we also give
results for double hard
contributions where the nuclear modification was omitted. Since we do not
know anything about the correct $x$-dependence of the higher twist matrix 
elements this gives an impression about the error we may at least assume for 
the higher twist calculation. 
The normalization constants for the twist-4 matrix elements
are chosen to be $\lambda^2 = 0.01$ GeV$^2$ and $C=0.005$ GeV$^2$.
In order to enable convenient comparison with cross sections we show all 
helicity amplitudes multiplied with the prefactor $N_\sigma=\alpha^2 / 
(64 \pi^3 S Q^2)$ from Eq.~(\ref{eq:dy}).

In Fig.~\ref{fig:wy} we give results for the helicity amplitudes $W_{\rm TL}$
and $W_\Delta$ as functions of rapidity. We observe that DH scattering
gives a large contribution at negative rapidities (the direction of the 
proton) which can easily balance 
the suppression of the twist-2 contribution by shadowing.
Soft hard scattering is strongly suppressed at these energies. 
$W_{\rm L}$ (not shown) qualitatively has the same behavior as $W_{\rm TL}$. 
The helicity flip amplitude $W_{\Delta}$ picks up only a small contribution 
from twist-2 and is entirely dominated by double hard scattering as already 
expected. This would be a good observable to pin down nuclear effects in
$p+A$ collision.
\begin{figure}[t]
\begin{center}
\includegraphics[width=7.0cm]{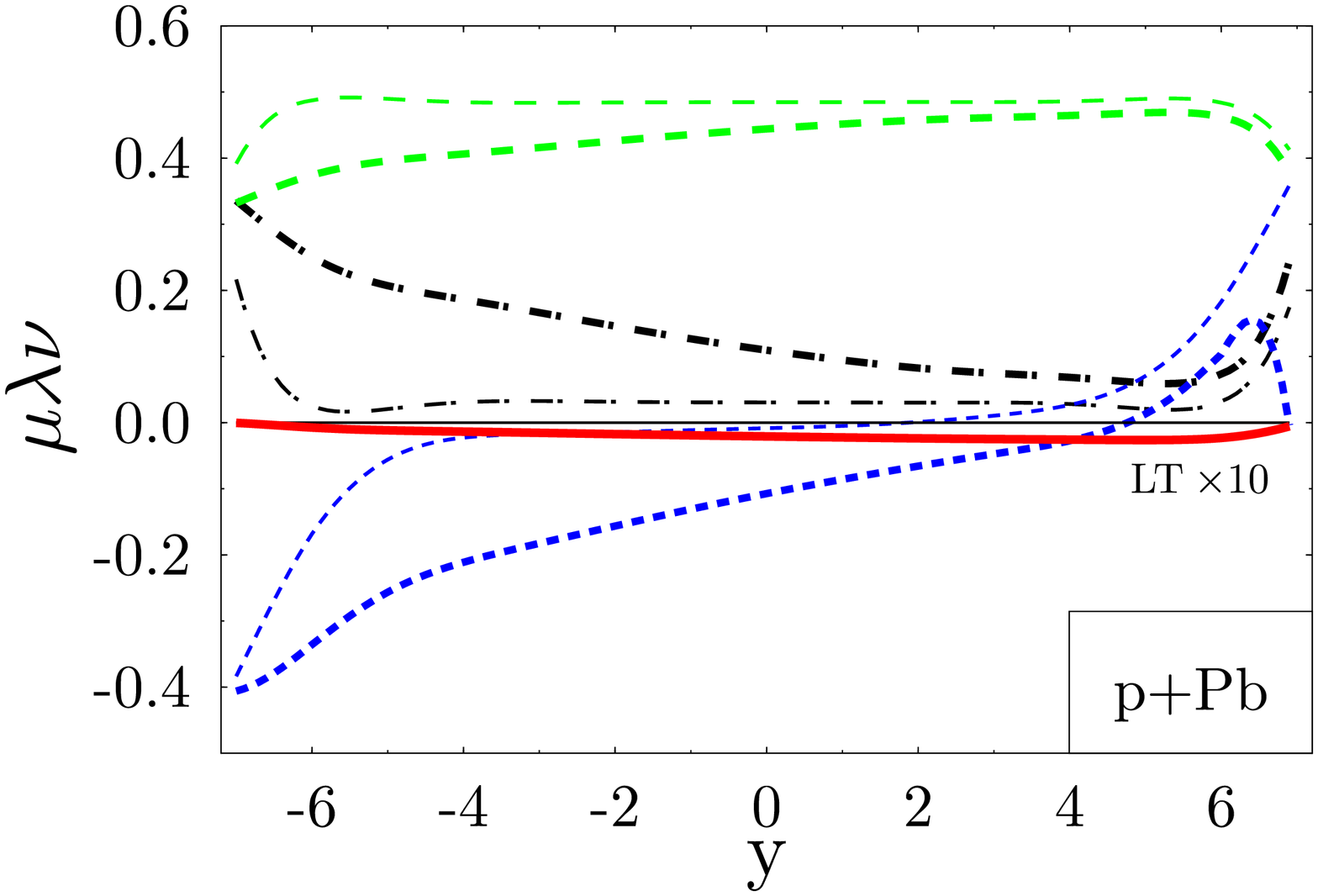} \hfill
\includegraphics[width=7.0cm]{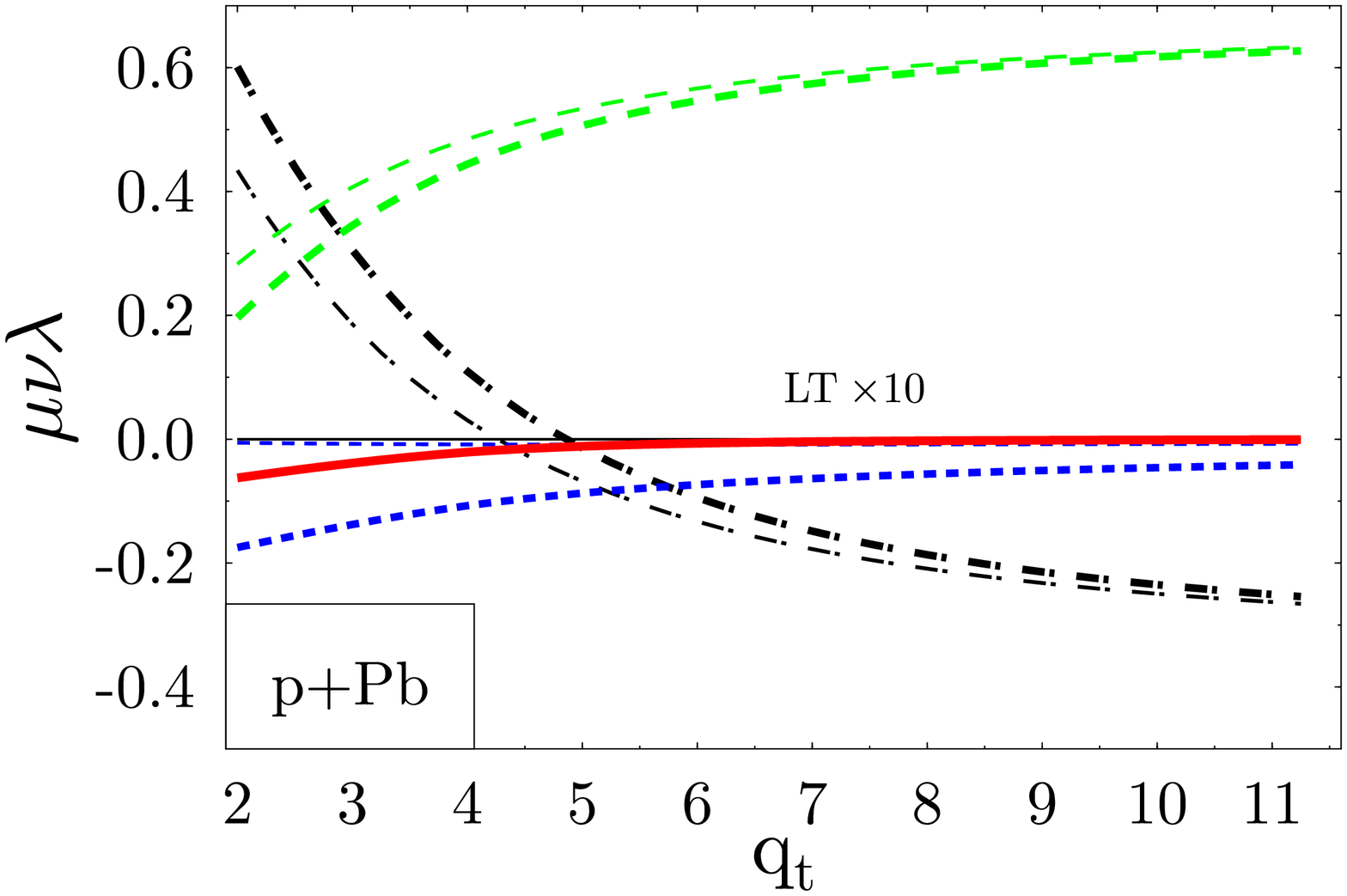}
\caption{The angular coefficients $\lambda$ (long dashed), $\mu$ 
 (short dashed) 
 and $\nu$ (dot dashed) for  $p+Pb$ at $\sqrt{S}=8.8$ TeV and $Q=5$ GeV as 
 functions of rapidity (left, at $q_\perp = 4$ GeV) and transverse momentum 
 (right,  at $y=0$). Thin lines represent pure twist-2 calculations, thick 
 lines show the results including twist-4. The solid line gives the violation of
 the Lam Tung relation $2\nu-(1-\lambda)$ scaled up by a factor of 10.}
 \label{fig:lmn}
\end{center}
\end{figure}
Fig.~\ref{fig:lmn} gives the full set of angular coefficients $\lambda$, 
$\mu$ and $\nu$ as functions of $y$ and $q_\perp$. Both the twist-2 
results and the modifications by twist-4 are given. The violation of the Lam 
Tung sum rule is numerically almost negligible since the soft hard 
contribution is so small. Note here that earlier experiments have discovered 
a large violation of the Lam Tung relation in $\pi+A$ collisions 
\cite{Conway:fs}. This issue is still not fully resolved. 

$p+A$ collisions offer the unique opportunity to study nuclear effects directly
via the rapidity dependence of observables. Helicity amplitudes and angular 
coefficients can help to pin down the role of nuclear enhanced higher twist.
Particularly promising are the helicity flip amplitude $W_{\Delta}$ and the 
coefficient $\mu$ which both vanish for
$p+p$. The usage of several different species of nuclei is advisable in order 
to address the important question of $A$ scaling.

\end{document}